\documentclass{SciPost}

\hypersetup{
    colorlinks,
    linkcolor={red!50!black},
    citecolor={blue!50!black},
    urlcolor={blue!80!black}
}

\usepackage{subcaption}
\usepackage[bitstream-charter]{mathdesign}
\DeclareSymbolFont{usualmathcal}{OMS}{cmsy}{m}{n}
\DeclareSymbolFontAlphabet{\mathcal}{usualmathcal}

\fancypagestyle{SPstyle}{
\fancyhf{}
\lhead{\colorbox{scipostdeepblue}{\bf \color{white} ~SciPost Physics Proceedings }}
\rhead{{\bf \color{scipostdeepblue} ~Submission }}

\fancyfoot[C]{\textbf{\thepage}}
}

\begin{document}

\pagestyle{SPstyle}

\begin{center}{\Large \textbf{\color{scipostdeepblue}{
Fair Universe Higgs Uncertainty Challenge\\
}}}\end{center}

\begin{center}\textbf{
Ragansu Chakkappai\textsuperscript{1,2$\star$},
Wahid Bhimji\textsuperscript{3},
Paolo Calafiura\textsuperscript{3},
Po-Wen Chang\textsuperscript{3},
Yuan-Tang Chou\textsuperscript{4},
Sascha Diefenbacher\textsuperscript{3},
Jordan Dudley\textsuperscript{5,3},
Steven Farrell\textsuperscript{3},
Aishik Ghosh\textsuperscript{6,3},
Isabelle Guyon\textsuperscript{2},
Chris Harris\textsuperscript{3},
Shih-Chieh Hsu\textsuperscript{4},
Elham E Khoda\textsuperscript{7,4,3},
Benjamin Nachman\textsuperscript{3},
Peter Nugent\textsuperscript{3},
David Rousseau\textsuperscript{1,2},
Benjamin Thorne\textsuperscript{3},
Ihsan Ullah\textsuperscript{2} and
Yulei Zhang\textsuperscript{4}
}\end{center}

\begin{center}
{\bf 1} Universit\'e Paris-Saclay, CNRS/IN2P3, IJCLab
\\
{\bf 2} ChaLearn
\\
{\bf 3} Lawrence Berkeley National Laboratory
\\
{\bf 4} University of Washington, Seattle
\\
{\bf 5} University of California, Berkeley
\\
{\bf 6} University of California, Irvine
\\
{\bf 7} University of California, San Diego
\\[\baselineskip]
$\star$ \href{mailto:email1}{\small fair-universe.lbl.gov}
\end{center}

\definecolor{palegray}{gray}{0.95}
\begin{center}
\colorbox{palegray}{
  \begin{tabular}{rr}
  \begin{minipage}{0.37\textwidth}
    \includegraphics[width=60mm]{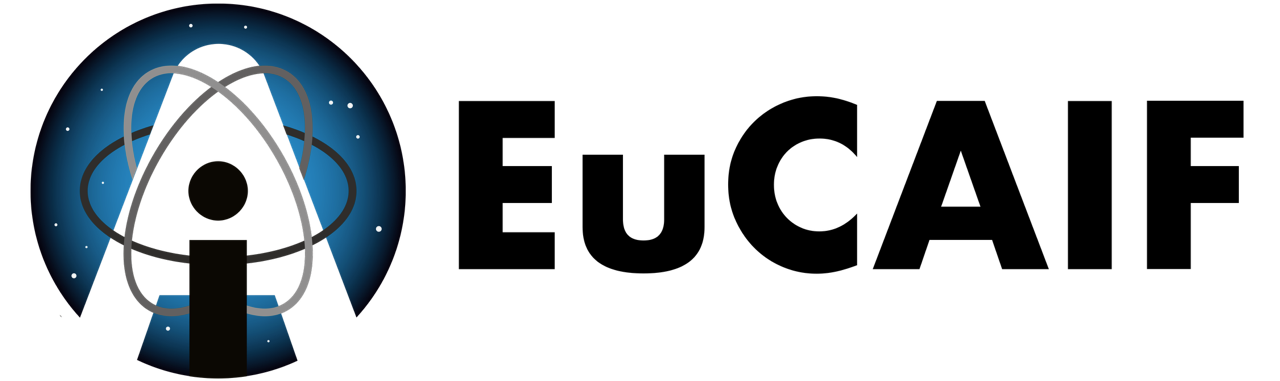}
  \end{minipage}
  &
  \begin{minipage}{0.5\textwidth}
    \vspace{5pt}
    \vspace{0.5\baselineskip} 
    \begin{center} \hspace{5pt}
    {\it The 2nd European AI for Fundamental \\Physics Conference (EuCAIFCon2025)} \\
    {\it Cagliari, Sardinia, 16-20 June 2025
    }
    \vspace{0.5\baselineskip} 
    \vspace{5pt}
    \end{center}
    
  \end{minipage}
\end{tabular}
}
\end{center}

\section*{\color{scipostdeepblue}{Abstract}}
\textbf{\boldmath{%
This competition in high-energy physics (HEP) and machine learning was the first to strongly emphasise uncertainties in $(H \rightarrow \tau^+ \tau^-)$ cross-section measurement. Participants were tasked with developing advanced analysis techniques capable of dealing with uncertainties in the input training data and providing credible confidence intervals. The accuracy of these intervals was evaluated using pseudo-experiments to assess correct coverage. The dataset is now published in Zenodo, and the winning submissions are fully documented.
}}

\vspace{\baselineskip}

\noindent\textcolor{white!90!black}{%
\fbox{\parbox{0.975\linewidth}{%
\textcolor{white!40!black}{\begin{tabular}{lr}%
  \begin{minipage}{0.6\textwidth}%
    {\small Copyright attribution to authors. \newline
    This work is a submission to SciPost Phys. Proc. \newline
    License information to appear upon publication. \newline
    Publication information to appear upon publication.}
  \end{minipage} & \begin{minipage}{0.4\textwidth}
    {\small Received Date \newline Accepted Date \newline Published Date}%
  \end{minipage}
\end{tabular}}
}}
}




\section{Introduction}

Ten years ago, part of our team co-organised the Higgs Boson Machine Learning Challenge (HiggsML~\cite{pmlr-v42-cowa14,kaggle_higgsml2014}. This challenge has significantly heightened interest in applying Machine Learning (ML) techniques within High-Energy Physics (HEP) and, conversely, has exposed physics issues to the ML community. However, the other challenge which remains, and \textit{must} be tackled for future discovery, is how to effectively quantify and reduce uncertainties, including understanding and controlling \textit{systematic} uncertainties. The traditional way to address this is to estimate the systematic uncertainty in the parameter estimation using shifted datasets and propagating that uncertainty to the final error prediction. However, this does not address the fundamental issue of biased ML models. In the past 10 years, advanced efforts that integrate uncertainties into the ML training include approaches that explicitly depend on nuisance parameters~\cite{Cranmer:2015bka,Baldi:2016fzo,Brehmer:2019xox,Brehmer:2018hga,Brehmer:2018kdj,Brehmer:2018eca,Nachman:2019dol,Ghosh:2021roe,Rozet:2021diu,ATLAS:2024ynn}, that are insensitive to nuisance parameters~\cite{Blance:2019ibf,Englert:2018cfo,Louppe:2016ylz,Dolen:2016kst,Moult:2017okx,Stevens:2013dya,Shimmin:2017mfk,Bradshaw:2019ipy,ATL-PHYS-PUB-2018-014,DiscoFever,Wunsch:2019qbo,Rogozhnikov:2014zea,10.1088/2632-2153/ab9023,clavijo2020adversarial,Kasieczka:2020pil,Kitouni:2020xgb,Estrade:2019gzk,Ghosh:2021hrh}, that use downstream test statistics in the initial training~\cite{Wunsch:2020iuh,CMS:2025cwy,Heinrich:2022qlq,Elwood:2020pik,Xia:2018kgd,DeCastro:2018psv,Charnock_2018,Alsing:2019dvb,Simpson:2022suz,Feichtinger:2021uff,Layer:2023lwi}, and that use Bayesian neural networks for estimating uncertainties~\cite{Kasieczka:2020vlh,Bollweg:2019skg,Araz:2021wqm,Bellagente:2021yyh}.  Many of these topics were covered in recent forward-looking review-type articles in Refs.~\cite{Dorigo:2020ldg, Chen:2022pzc}. Unfortunately, many of these works are often published with different datasets and problem settings, making comparing between methods a challenge. This motivated the creation of a publicly available challenge with a large dataset focused on uncertainty quantification. The Fair Universe Uncertainty Challenge~\cite{bhimji2025fair} was accepted as a NeurIPS 2024 Challenge, and the paper is accepted in the Dataset and Benchmark track of NeurIPS 2025. 

\section{Challenge Setting}

\begin{figure}
\begin{center}
\includegraphics[width=6cm]{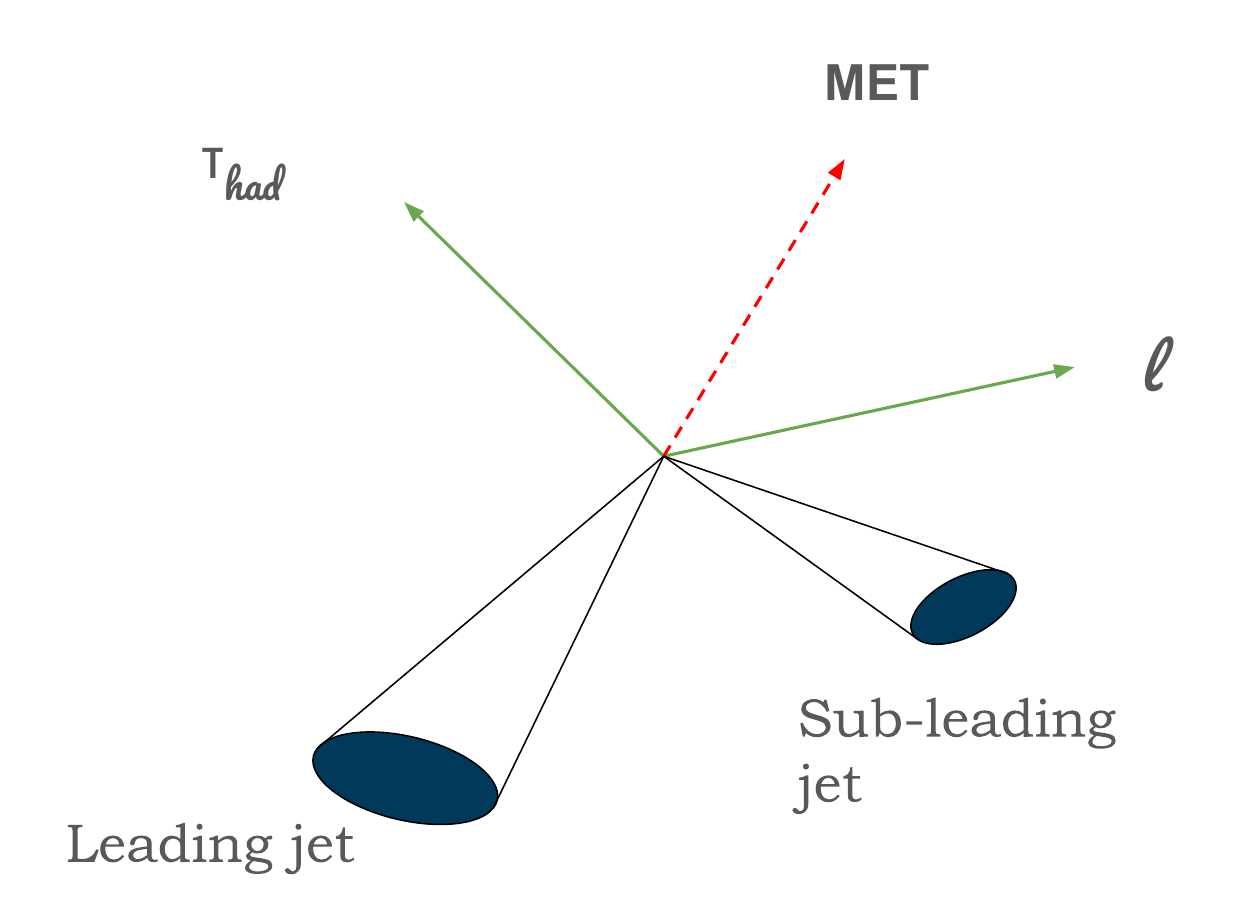}
\end{center}
\caption{Diagram of the particles in the final state chosen: one lepton, one tau hadron, up to two jets, and the missing transverse momentum vector\cite{bhimji2025fair}}
\label{fig_particles_diag}
\end{figure}

The participant's objective is to develop an algorithm to estimate the amount of Higgs boson signal and provide a 1 $\sigma$ confidence interval to that prediction. The physics process in question is the Higgs boson decaying into two $\tau$ particles: $(H \rightarrow \tau^+ \tau^-)$ (see \autoref{fig_particles_diag}). The parameter to be estimated is the signal strength $\mu$, which is defined as the ratio of the observed number of signal events to the expected number of signal events in the standard model. The main background in the challenge is $(Z \rightarrow \tau^+ \tau^-)$ events. These events are a thousand times more likely to be produced than the Higgs Boson. As this challenge focuses on uncertainties, the participant's model will be tested on a shifted dataset that would have systematics with unknown values of nuisance parameters. Further, to correctly evaluate the confidence interval (CI) given by the participants, we test the participant's model several times (10 trials of 100 pseudo-experiments in the Public phase and 1000 trials of 100 pseudo-experiments in the Private phase), each trial with a given value of signal strength $\mu$ randomised between 0.1 and 3.

\section{Datasets and Systematics} \label{sec:datasets}
The challenge dataset was generated using the Pythia8~\cite{Sjostrand:2007gs} event generator in conjunction with the Delphes 3.5~\cite{deFavereau:2013fsa} detector simulator. 
The Dataset~\cite{https://doi.org/10.5281/zenodo.15131565} aimed to be at least 200 times larger than the equivalent number of events in the LHC. The dataset is in a tabular form with 28 high-level variables, 16 primary variables ($p_T$,$\eta$,$\phi$) of $\tau_{\text{lep}}$, $\tau_{\text{had}}$ and jets and 12 derived variables. We provided a shifting function to transform the datasets for a given set of 6 different nuisance parameter values, three feature-distorting systematics ( Tau-hadron Energy Scale (TES) Jet Energy Scale (JES) Soft Missing Transverse Energy  (Soft MET)), which change the values of different features in the datasets and three normalisation systematics, which change the numbers of each background event-category or weights (Total Background Normalisation, Di-boson Background Normalisation, $t\bar{t}$ Background Normalisation). 

\section{Evaluation and Scoring}

\begin{figure}[t]
    \centering
    \begin{subfigure}[b]{0.40\textwidth}
    \includegraphics[width=8cm]{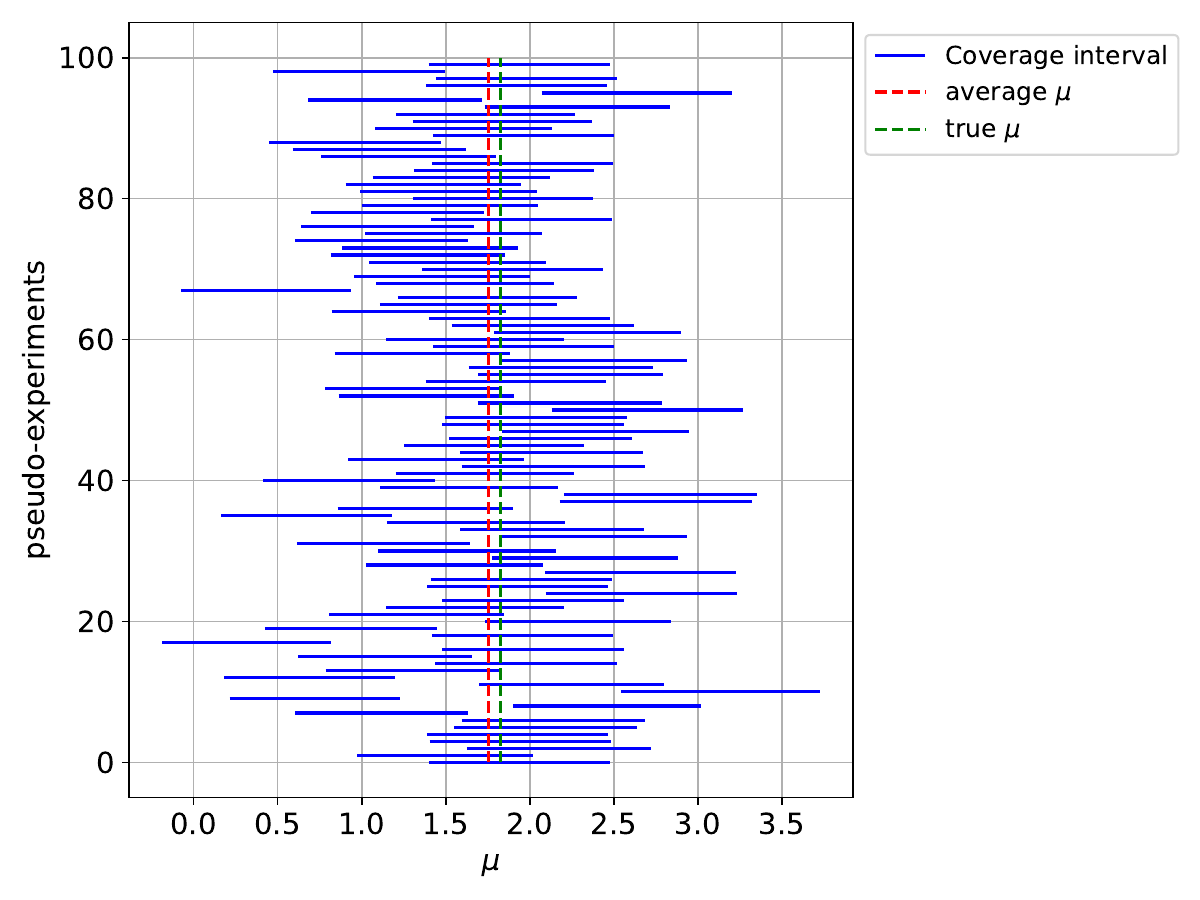}
    \caption{}
    \label{fig:mu_distribution}
    \end{subfigure}
    \centering
    \begin{subfigure}[b]{0.40\textwidth}
    \includegraphics[width=6cm]{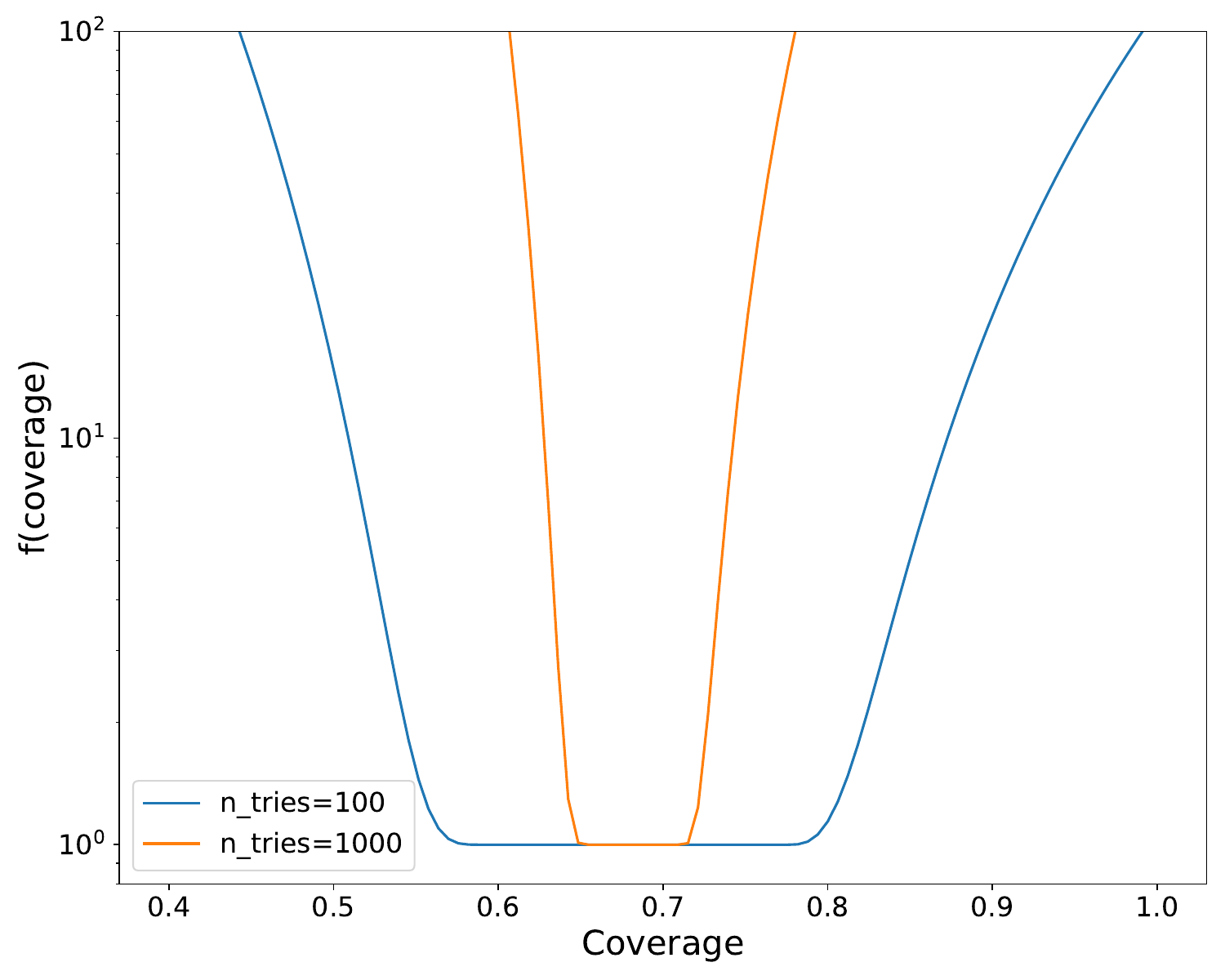}
    \caption{}
    \label{fig:coverage_plot}
    \end{subfigure}   
    \caption{ (\ref{fig:mu_distribution})\textit{Coverage plot}: predicted intervals (blue lines) for each pseudo experiment generated for a given $\mu_{\rm true}$ (vertical dotted line). The coverage  (here $70\pm5\%$) is determined by the fraction of horizontal blue lines intersected by the vertical line. The average width of the interval is here 1.068. (\ref{fig:coverage_plot}) \textit{Coverage penality}: 1D function to penalise models with poor coverage.  \cite{bhimji2025fair}} 
    \label{fig_coverage}
\end{figure}

The scoring algorithm evaluates the coverage of the quoted CI by checking the percentage of times where the true $\mu$ (The green vertical line in \autoref{fig:mu_distribution}) falls within the quoted CI (The blue horizontal lines in \autoref{fig:mu_distribution}). Ideally, the coverage should be 68.27\%. Since the number of pseudo-experiments is limited, the coverage can fluctuate. To properly account for this, we designed a special coverage penalty function ($f(x)$)\autoref{fig:coverage_plot} which gives 1 when the coverage is near 68.27\%  and a much higher value if the model is overconfident or underconfident. The final score is the negative log of the mean width of CI times the coverage function $f(c)$. This means to get a high score, one must minimise the CI without sacrificing the coverage. 

\section{Competition results and best submissions}
\label{sec:results}
\begin{figure}[tb]
    \centering
    \begin{subfigure}[b]{0.32\textwidth}
        \includegraphics[width=\textwidth]{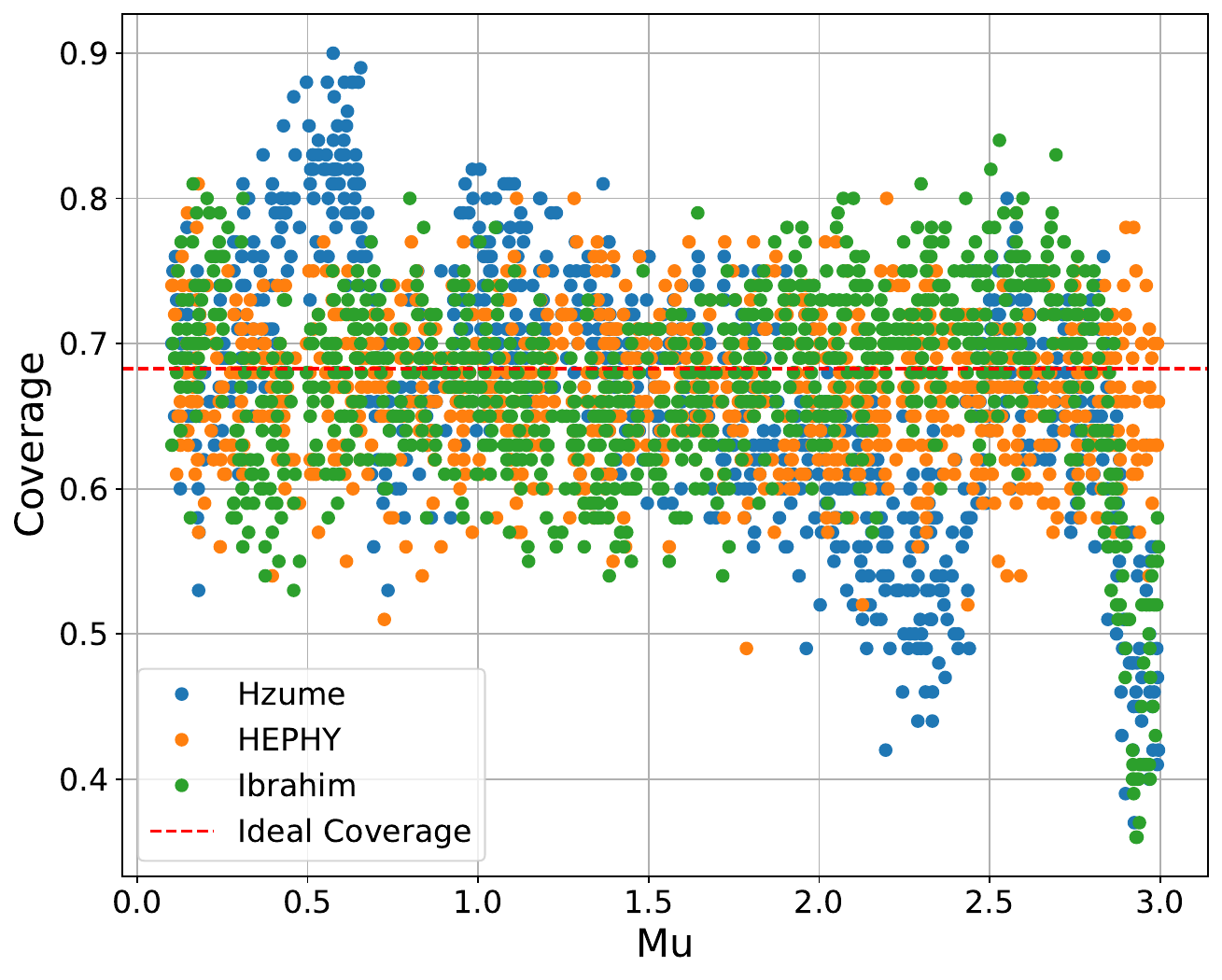}
        \caption{}
        \label{fig:coverage_vs_mu}
    \end{subfigure}
    \hfill 
    \begin{subfigure}[b]{0.32\textwidth}
        \includegraphics[width=\textwidth]{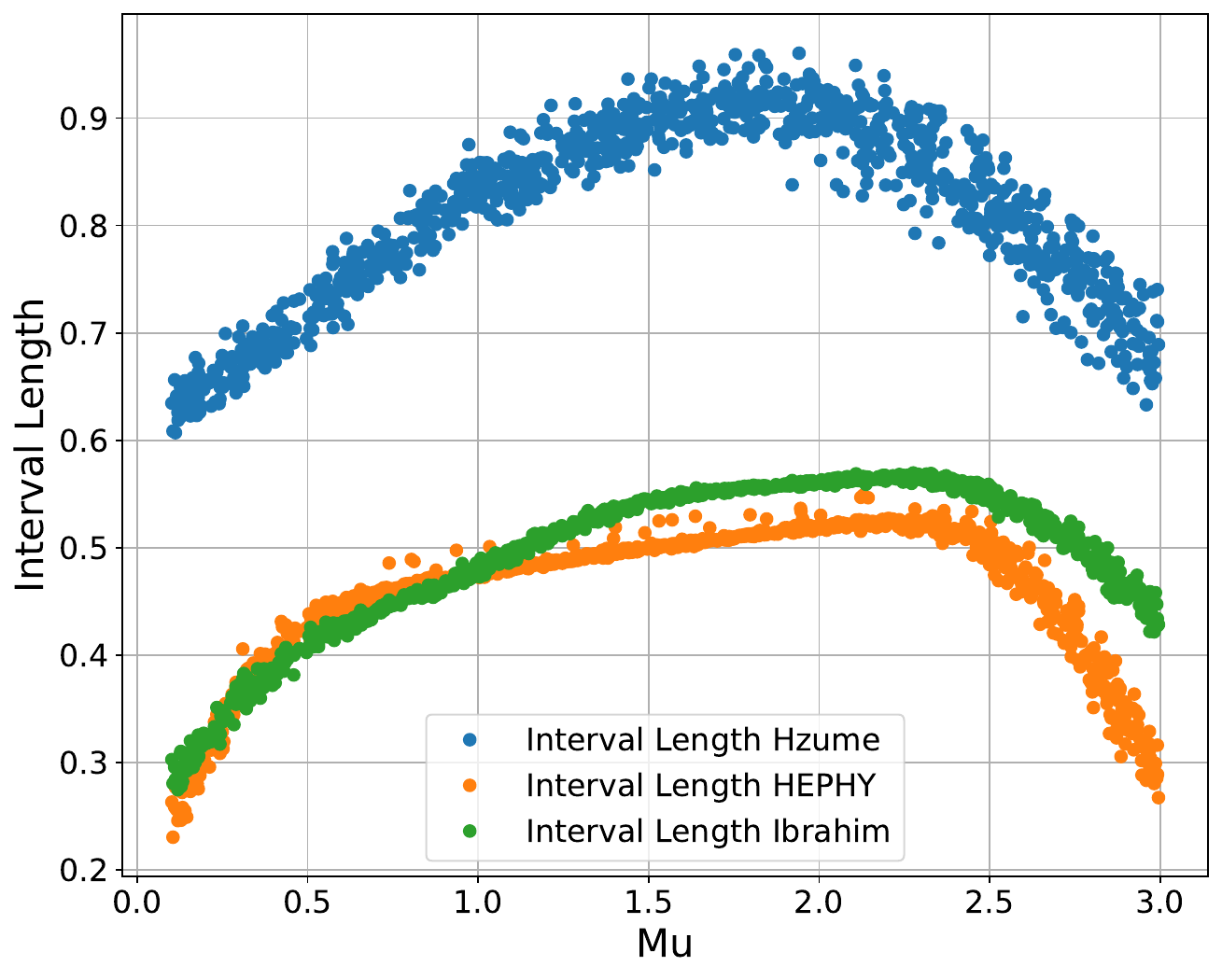}
        \caption{}
        \label{fig:interval_length_vs_mu}
    \end{subfigure}
    \hfill
    \begin{subfigure}[b]{0.32\textwidth}
        \includegraphics[width=\textwidth]{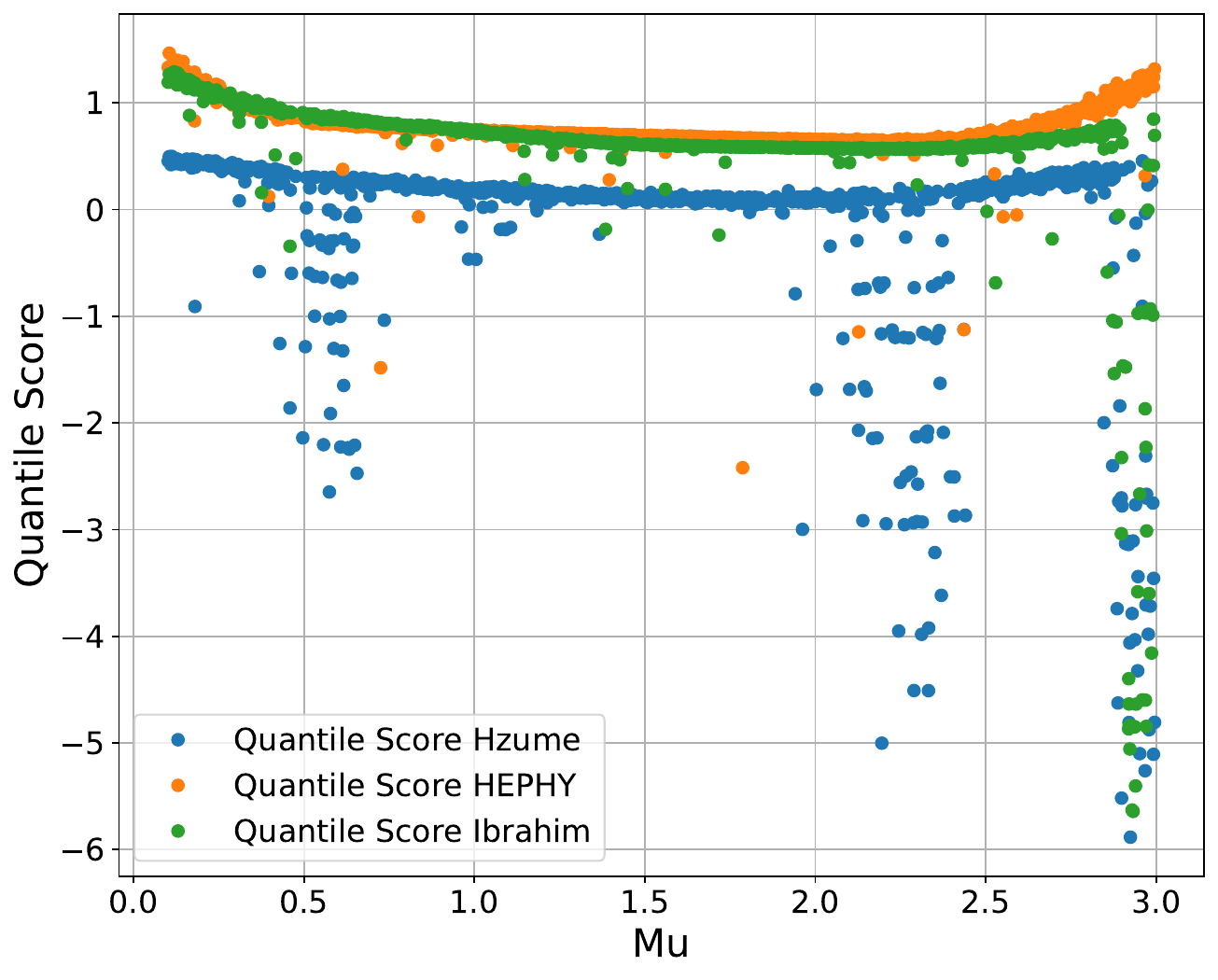}
        \caption{}
        \label{fig:quantile_score_vs_mu}
    \end{subfigure}
    \caption{Comparative study of the three finalists (blue for Hzume, orange for HEPHY and green for Ibrahim's model) with 1000 trials of 100 pseudo-experiments. (\ref{fig:coverage_vs_mu}) the coverage from each trial, (\ref{fig:interval_length_vs_mu}) the average CI width  and (\ref{fig:quantile_score_vs_mu}) the quantile score.  \cite{bhimji2025fair}}
    \label{fig:three_scores}
\end{figure}

At the end of the Public phase, a clear trio was at the top of the public leaderboard: HEPHY with a quantile score of 0.878, followed by Ibrahime (0.823) and Hzume (0.179).  All submissions have been reevaluated on a new dataset (i.i.d. to the original one). All submissions were run on the same pseudo-experiments. 
\autoref{fig:three_scores} shows the results for all trials for the trio. 
In the final phase, scores HEPHY and IBRAHIME were very close; hence, additional bootstrap analysis of the variance of these results showed that submissions of HEPHY and IBRAHIME cannot be reliably ranked, hence the final rankings :

\begin{itemize}
\item 1st tie: HEPHY with \textit{"Unbinned inclusive cross-section measurements with machine-learned systematic uncertainties"}~\cite{benato2025unbinnedinclusivecrosssectionmeasurements} (Lisa Benato, Cristina Giordano, Claudius Krause, Ang Li,  Robert Schöfbeck, Maryam Shooshtari, Dennis Schwarz, Daohan Wang) from Vienna’s Institute of High Energy Physics (HEPHY) in Austria wins \$2000.
\item 1st tie IBRAHIME (Ibhrahim Elsharkawy) with \textit{"Contrastive Normalizing Flows for Uncertainty-Aware Parameter Estimation"}~\cite{elsharkawy2025contrastivenormalizingflowsuncertaintyaware} (Ibrahim Elsharkawy) from University of Illinois at Urbana-Champaign, USA wins \$2000.
\item  3rd HZUME (Hashizume Yota) with \textit{"Decision-Tree Aggregated Features and Hybrid Bin-Classifier/Quantile-Regressor"}  from Kyoto University, Japan wins \$500
\end{itemize}

\section*{Conclusion}

The competition brought together cutting-edge infrastructure for AI training and inference with large datasets and a standardised scoring for uncertainty computation. The dataset is permanently stored in Zenodo and is publicly available, which ensures its possible use as a standard benchmark for uncertainty quantification in HEP. The competition concluded with 2 competitive yet different winning solutions from HEPHY and IBRAHIME, suggesting the possibility of combining these models. We believe that submissions from the FAIR Universe challenge can push the boundaries of Uncertainty-Aware Artificial Intelligence in the coming years, within and outside the HEP community. 

\section*{Acknowledgements}
We are grateful to the US Department of Energy, Office of High Energy Physics, and the subprogram on Computational High Energy Physics, for sponsoring this research, as well as to the ANR Chair of Artificial Intelligence HUMANIA (ANR-19-CHIA-0022). Seminal discussions contributing to this work took place at the workshop “Artificial Intelligence and the Uncertainty Challenge in Fundamental Physics,” sponsored by the CNRS AISSAI Center and the DATAIA Institute (ANR-17-CONV-003), and hosted at Institut Pascal (ANR-11-IDEX-0003-01) at Université Paris-Saclay. This research used resources of the National Energy Research Scientific Computing Center (NERSC), a Department of Energy Office of Science User Facility using NERSC award HEP-ERCAP0032917.

\bibliography{ref.bib}


\end{document}